\documentclass[10pt]{article}
\usepackage{tabularx}
\usepackage{booktabs}
\usepackage{graphicx}
\usepackage{array} 
\usepackage{listings}
\usepackage{xcolor}
\usepackage{amssymb}
\usepackage{amsmath}
\usepackage{float}
\usepackage[round,authoryear]{natbib} 
\usepackage[margin=1in]{geometry} 
\usepackage{setspace} 
\usepackage{hyperref}
\hypersetup{
    colorlinks=true,
    citecolor=blue,
    linkcolor=blue,
    urlcolor=blue
}
\newcolumntype{Y}{>{\centering\arraybackslash}X}

\lstdefinestyle{mystyle}{
    backgroundcolor=\color{gray!10},   
    commentstyle=\color{green},
    keywordstyle=\color{blue},
    numberstyle=\tiny\color{gray},
    stringstyle=\color{red},
    basicstyle=\ttfamily\footnotesize, 
    breakatwhitespace=false,         
    breaklines=true,                 
    captionpos=b,                    
    keepspaces=true,                 
    numbers=left,                    
    numbersep=5pt,                  
    showspaces=false,                
    showstringspaces=false,
    showtabs=false,                  
    tabsize=2
}

\begin{document}
\title{Evaluation of ChatGPT's Smart Contract Auditing Capabilities Based on Chain of Thought}
\author{Yuying Du \\ Salus Security \and Xueyan Tang \\Salus Security \thanks{Corresponding author. Email:77728@gmail.com}}
\date{\today}
\maketitle

\begin{abstract}
Smart contracts, as a key component of blockchain technology, play a crucial role in ensuring the automation of transactions and adherence to protocol rules. However, smart contracts are susceptible to security vulnerabilities, which, if exploited, can lead to significant asset losses. In recent years, numerous incidents of smart contract attacks have highlighted the importance of smart contract security audits. This study explores the potential of enhancing smart contract security audits using the GPT-4 model. We utilized a dataset of 35 smart contracts from the SolidiFI-benchmark vulnerability library, containing 732 vulnerabilities, and compared it with five other vulnerability detection tools to evaluate GPT-4's ability to identify seven common types of vulnerabilities. Moreover, we assessed GPT-4's performance in code parsing and vulnerability capture by simulating a professional auditor's auditing process using CoT(Chain of Thought) prompts based on the audit reports of eight groups of smart contracts. We also evaluated GPT-4's ability to write Solidity Proof of Concepts (PoCs). Through experimentation, we found that GPT-4 performed poorly in detecting smart contract vulnerabilities, with a high Precision of 96.6\%, but a low Recall of 37.8\%, and an F1-score of 41.1\%, indicating a tendency to miss vulnerabilities during detection. Meanwhile, it demonstrated good contract code parsing capabilities, with an average comprehensive score of 6.5, capable of identifying the background information and functional relationships of smart contracts; in 60\% of the cases, it could write usable PoCs, suggesting GPT-4 has significant potential application in PoC writing. These experimental results indicate that GPT-4 lacks the ability to detect smart contract vulnerabilities effectively, but its performance in contract code parsing and PoC writing demonstrates its significant potential as an auxiliary tool in enhancing the efficiency and effectiveness of smart contract security audits.
\end{abstract}



\section{Introduction}
Smart contracts are automated programs designed to execute the necessary actions stipulated in agreements or contracts automatically. Once completed, transactions are traceable and irreversible. They focus on setting and deploying rules on the network to complete transactions. Ethereum is a blockchain platform that enables developers to create and deploy smart contracts \cite{ethereum}. Essentially, smart contracts are built based on algorithmic and programming choices to ensure correct execution \cite{vul_audit_intro}. Due to their code-based and immutable deployment nature, errors in the code can lead to vulnerabilities in smart contracts, making them susceptible to attacks and nearly impossible to modify or remove after deployment \cite{vul_audit2_intro}. Vulnerable smart contracts can often lead to more severe issues. For instance, in February 2022, Wormhole was robbed of \$325 million when a hacker exploited the smart contract on the SOL-ETH bridge to cash out without depositing any collateral \cite{cointg}. In March 2023, Euler Finance lost \$197 million due to a smart contract vulnerability \cite{br}. On July 30, 2023, Curve suffered a loss of about \$70 million due to vulnerabilities in the Vyper language it used, highlighting the necessity of comprehensive smart contract audits \cite{curve}. These incidents of smart contract attacks demonstrate the unpredictable financial losses that can occur if smart contracts, which control a significant amount of cryptocurrency and financial assets, are targeted and compromised.

Smart contract security auditing \cite{de2021impact} is the process of assessing the security and reliability of smart contracts. Smart contract audits involve detailed analysis of the contract code to identify security issues, incorrect and inefficient coding, and determine solutions for these problems. Conducting thorough audits helps minimize the risks and potential losses that could result from security vulnerabilities after deploying smart contracts. In practice, smart contracts often present more complex transaction scenarios or very intricate execution logic, significantly increasing the burden of auditing work \cite{complex_contracts_intro}. As of 2022, a variety of smart contract vulnerability analysis tools, including both static and dynamic analysis tools, totaling approximately 86, have been developed \cite{audit_tools_intro}. The number of related analysis tools continues to grow.

In recent years, Large Language Models (LLMs) \cite{chang2023survey} have been in a phase of iteration and rapid development, and their outstanding performance in generation and analysis tasks has become a transformative force in various fields. The current state-of-the-art LLM, Generative Pre-trained Transformer (GPT) \cite{wubrief}, has evolved to GPT-4 \cite{chatgpt4}, showcasing exceptional performance in text analysis and generation. In the context of smart contract audits, GPT-4 has the potential to serve as a powerful analytical tool. To better leverage GPT-4 in smart contract auditing tasks, it's worth assessing its capability in analyzing smart contracts and detecting vulnerabilities.

This paper evaluates the capacity of GPT-4 in the domain of smart contract audits, with the knowledge cut-off for the GPT-4 used in this experiment being April 2023. Our main contributions are as follows:

1. To assess GPT-4's capability in detecting smart contract vulnerabilities, we chose the SolidiFI-benchmark \cite{ghaleb2020effective} vulnerability library as our dataset, selecting 35 smart contracts for testing, which collectively contain 732 injected vulnerabilities, to evaluate GPT-4's detection performance across 7 common types of vulnerabilities.

2. To simulate real-world audit scenarios, we conducted experiments based on audit reports of 8 sets of smart contracts, which revealed a total of 60 vulnerabilities, covering 18 types of audit vulnerabilities. We designed a prompt based on CoT(Chain of Thought) \cite{CoT} reasoning. The goal of this prompt is to require GPT-4 to mimic the way a professional auditor would audit a smart contract, to assess GPT-4's code parsing capabilities and its ability to identify vulnerabilities within.

3. To evaluate GPT-4's capability in identifying and verifying security vulnerabilities, we also selected 10 smart contracts and adopted a dual-modal experimental design to test GPT-4's ability in writing PoC(Proof of Concept ) in Solidity.

This paper primarily focuses on GPT-4's ability in vulnerability detection, code parsing, and PoC writing for smart contracts. The details of our focus are summarized in the table\ref{table1} below:

\begin{table}[ht]
\centering
\begin{tabular}{|p{4cm}|p{8cm}|}
\hline
\textbf{Evaluation Aspect} & \textbf{Description} \\
\hline
Vulnerability Detection & Evaluate the efficiency and accuracy of GPT-4 in detecting vulnerabilities in smart contracts. This involves analyzing its capability to identify security issues such as reentrancy attacks, overflow, etc. \\
\hline
Code Parsing Capability & Test whether GPT-4 can accurately understand and analyze the business logic, context, and code structure of smart contracts. This includes understanding the functionality of the contract, the relationships between function calls, etc. \\
\hline
PoC Writing Ability & Examine GPT-4's ability to write PoCs, assessing its capability to recognize and validate potential security vulnerabilities. \\
\hline
\end{tabular}
\caption{GPT4 evaluation aspect}
\label{table1}
\end{table}

Through these experiments, we can more comprehensively assess the role of GPT-4 in auditing tasks, which is beneficial for us to better utilize GPT-4 to assist auditors in their work and improve efficiency. This not only has the potential to reduce manpower costs in the auditing process but also to enhance the quality and accuracy of audits. More importantly, such evaluations help us understand and develop the application of artificial intelligence in the field of smart contract security, providing references for the technological development and security strategy formulation in the field of cybersecurity. For your convenience, our experimental data is here: https://github.com/Mirror-Tang/Evaluation-of-ChatGPT-s-Smart-Contract-Auditing-Capabilities-Based-on-Chain-of-Thought/tree/master

\section{Related Work}
In this section, we will introduce the related work in smart contract audits, mainly focusing on vulnerability detection methods, smart contract audit work based on GPT-4, and Proof of Concept.

\subsection{Vulnerability detection methods}
Smart contract auditing \cite{de2021impact} involves a detailed analysis of a protocol's smart contract code to identify security vulnerabilities, poor coding practices, and inefficient code, followed by proposing solutions to these issues. During the smart contract auditing process, some automated tools are also used.

Several static auditing tools based on Solidity code have achieved significant performance in detecting vulnerabilities in smart contracts. Slither \cite{feist2019slither} uses the Abstract Syntax Tree (AST) to analyze Solidity code, thereby detecting vulnerabilities and code flaws in smart contracts. SmartCheck \cite{tikhomirov2018smartcheck} improves contract security by setting various rules to detect different types of vulnerabilities. Securify is a security scanner for Ethereum smart contracts supported by the Ethereum Foundation and ChainSecurity, supporting 37 types of vulnerabilities \cite{securify}.

Symbolic execution tools have also demonstrated significant performance in smart contract auditing. Mythril \cite{ruskin1980mythril} combines symbolic execution, taint analysis, and control flow analysis to review the security of Ethereum smart contracts. This approach allows Mythril to analyze not only Solidity source code but also compiled EVM bytecode. Manticore \cite{manticore} is a symbolic execution tool used for analyzing smart contracts and binary files. Oyente \cite{oyente} is a symbolic execution-based smart contract analysis tool that supports the detection of reentrancy, unhandled exceptions, transaction-ordering dependencies, and timestamp dependencies.

Although static analysis tools and symbolic execution tools provide some level of assistance in smart contract auditing work, this help is limited \cite{zhang2023demystifying}. These tools may not accurately identify all potential vulnerabilities and security risks, especially those involving complex business logic or specific blockchain environments. Static analysis tools can comprehensively check the code but often produce a high false-positive rate and struggle to fully inspect complex business logic \cite{statict}. On the other hand, symbolic execution tools, while testing contract behavior through program analysis, may not cover all execution paths and are resource-intensive, especially when dealing with complex and advanced security issues \cite{9230065}.

\subsection{GPT-4-based Smart Contract Audit}
ChatGPT is a large language model-based chatting robot developed by OpenAI \cite{openai2023}. The core of ChatGPT is Artificial Intelligence Generated Content (AIGC)-related task, and the critical technologies underlies ChatGPT consist of Pre-trained Language Model, In-context Learning and Reinforcement Learning from Human Feedback  \cite{background_chatgpt}. ChatGPT is the extension of GPT series, which mainly includes GPT-1, GPT-2, GPT-3, GPT-3.5 and GPT-4. The most widely-used models currently is GPT-4. Pre-traning data size and learning target for each GPT model vary. In terms of pre-training data size, GPT-1 trained on around 5 GB of data, and GPT-2 trained on 40 GB. GPT-3 achieves 45 TB of data in pre-training phase, while the data size of training GPT-4 is not published. For learning target, unsupervised learning, multi-task learning, in-context learning and multi-modal learning are implemented respectively in GPT-1, GPT-2, GPT-3 and GPT-4 \cite{background_chatgpt}. Advances in learning objectives and increasing amounts of training data make the newly released GPT model, GPT-4, powerful.

The current latest updated GPT model, GPT-4, performs well on text analysis and generation tasks, providing the expected potential for smart contract auditing. Multiple researchers conduct researches to evaluate the capabilities of GPT in smart contract audit. \cite{gpt_test1} focus on investigating the accuracy of GPT-4 in detecting vulnerabilities and it achieves 78.7\% true positive rate. \cite{sun2023gpt} propose GPTScan for smart contract logic vulnerability detection and precision achieves over 90\%. \cite{gpt_test2} provide systematic analysis of GPT-4 and propose an adversarial framework called GPTLENS for detecting a broad spectrum of vulnerabilities. \cite{chatgpt_pre} present an empirical study on the performances of ChatGPT vulnerabilities detection. 

In our experiments, we expanded the scope of evaluation, mainly assessing GPT-4's (with the knowledge cut-off in April 2023) ability to detect vulnerabilities in smart contracts, parse code, and its capability for PoC. This evaluation framework is more representative and comprehensive.

\subsection{Proof of Concept}
PoC vulnerability exploitation is a harmless attack on a computer or network. PoC demonstrates the feasibility and viability of an idea, and in the context of smart contract auditing, it is used to verify the effectiveness of vulnerabilities \cite{substack}. Writing PoCs for smart contracts is a key step in verifying their functionality, security, and performance. Through PoCs, developers can identify and verify potential security vulnerabilities in a controlled environment, evaluate the performance of the contract in processing transactions, including gas fees and execution efficiency. It is also used to verify the accuracy and consistency of contract logic, test interoperability with other contracts or external data sources, and more.

In terms of writing PoCs for smart contracts, Wei and other researchers have proposed a model called SemFuzz \cite{semfuzz}, which is a semantic-based system for auto-generating PoC vulnerabilities. SemFuzz utilizes error reports of software to automatically generate PoCs, which helps in more effectively identifying and fixing security vulnerabilities. Tools like Foundry \cite{foundry} and Hardhat \cite{hardhat} are widely used. Foundry provides a guide for writing PoC tests \cite{foundryPoC}, while Hardhat is also used for developing PoCs to verify vulnerabilities in smart contracts \cite{hardhatPoC}. The use of these tools demonstrates the importance and effectiveness of automation tools in the field of smart contract security. DefiHackLabs \cite{defihacklabs} provides a set of guidelines for audit steps \cite{substack}, an important part of which is about writing PoCs. This indicates that writing PoCs is not only a crucial step in identifying and verifying potential vulnerabilities during the smart contract audit process but also a key aspect in understanding and analyzing contract security.

\section{Methodology}

\subsection{Prompt Design}
According to some experiments on smart contracts conducted by  \cite{sun2023gpt}, \cite{gpt_test2}, and others using GPT, we found that the effectiveness of prompts significantly affects GPT-4's responses. Therefore, for the experiments in this paper, we designed clear and effective prompts, incorporating role definition, CoT, requirement specifications, and response format in our prompt design.

\subsubsection{Vulnerability Detection in the 
SolidiFI-benchmark Dataset}
This paper tests GPT-4's capability in identifying vulnerabilities in smart contracts, selecting 35 smart contracts with artificially injected vulnerabilities from the SolidiFI-benchmark dataset, covering 7 common types of vulnerabilities. These types include overflow/underflow, reentrancy, TOD, timestamp dependency, unchecked send, unhandled exceptions, and tx.origin vulnerabilities.

\begin{figure}[ht]
\centering 
\includegraphics[width=0.8\textwidth]{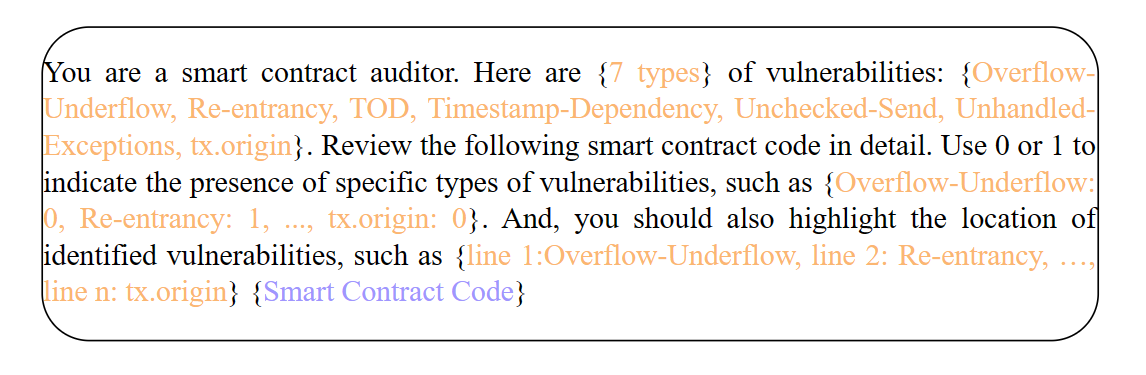} 
\caption{Vulnerability detecting prompt} 
\label{fig1} 
\end{figure}

We defined prompts as scenarios that fulfill specific roles and needs. As shown in Figure\ref{fig1}, we provided clear role definitions, prior knowledge, and response formats in our prompts. We tasked GPT-4 with acting as a smart contract auditor, to identify the seven types of vulnerabilities present in the contracts we provided, and to specify their exact locations.

\subsubsection{Smart Contract Code Analysis}
We examined vulnerabilities in 8 sets of audit reports, covering 18 types of smart contract vulnerabilities, including Business Logic, Access Control, Data Validation, Numerics, Reentrancy, Cryptography, Denial of Service, Upgradeable, and others.

To better assess GPT-4's capabilities in detecting vulnerabilities and parsing code in smart contracts, we evaluated its detection results against audit reports as a benchmark. We designed a prompt based on the CoT approach. The aim of this prompt was to require GPT-4 to mimic the manner in which auditors audit smart contracts. The entire audit process involves thoroughly understanding the background information and objectives of the smart contract, as well as conducting a comprehensive analysis of the logic behind its functions and call relationships. Based on the aforementioned procedure, we further requested GPT-4 to identify potential vulnerabilities.
\begin{figure}[ht]
\centering 
\includegraphics[width=0.8\textwidth]{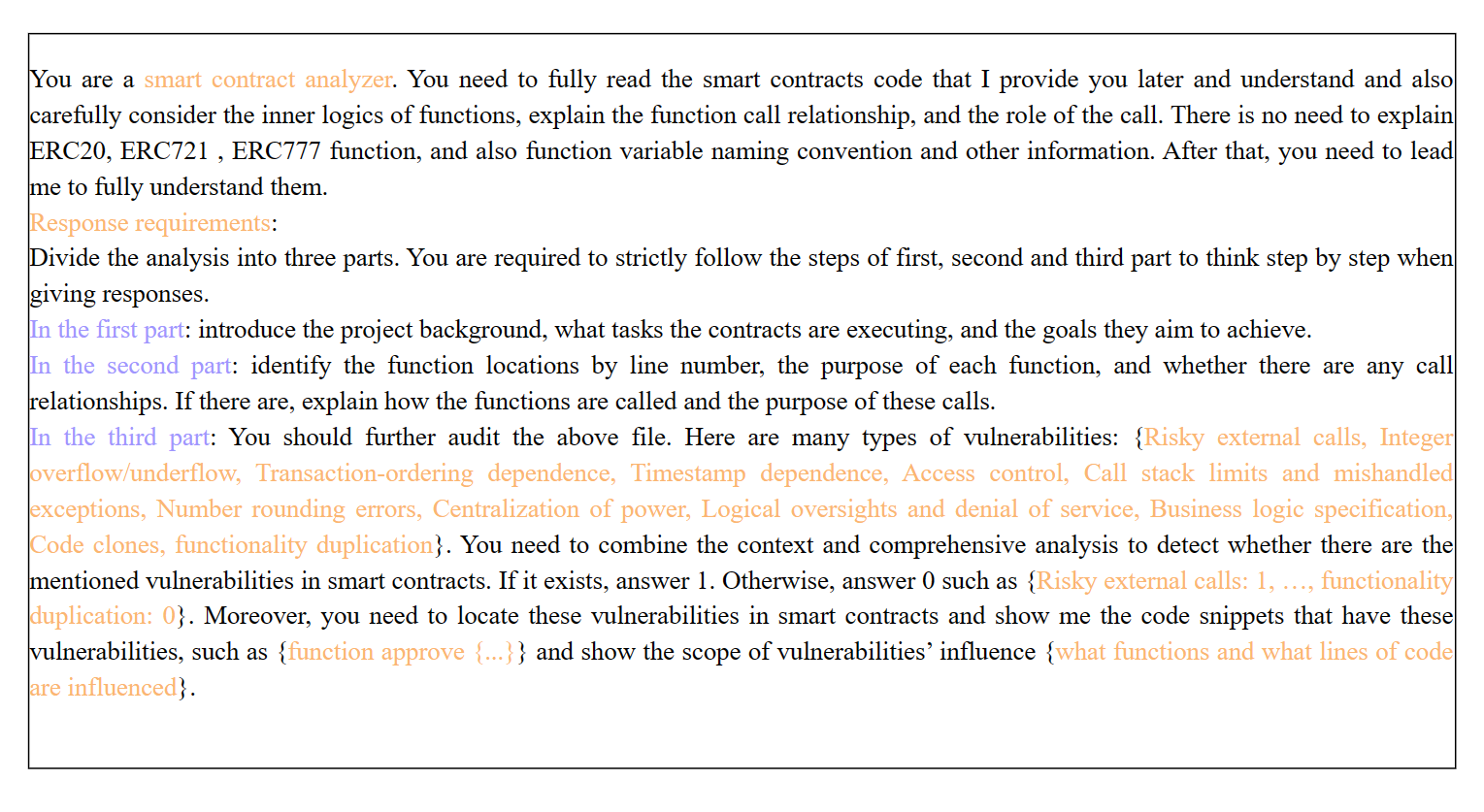} 
\caption{Smart contract code detecting prompt} 
\label{Figure2} 
\end{figure}

The prompt contains three critical thinking steps, which corresponds to the above critical procedures for manual audit. We combine role definition, chain of thought, requirement specification and response format in the prompt design Figure\ref{Figure2}. We also excluded certain functions that did not require specific explanations, such as ERC20, ERC721, and ERC777 functions, as well as information on function variable naming conventions. This is because such content is assumed to be well-known to auditors and does not require explanation.

\subsubsection{PoC Quality Assessment}
\begin{figure}[ht]
\centering 
\includegraphics[width=0.6\textwidth]{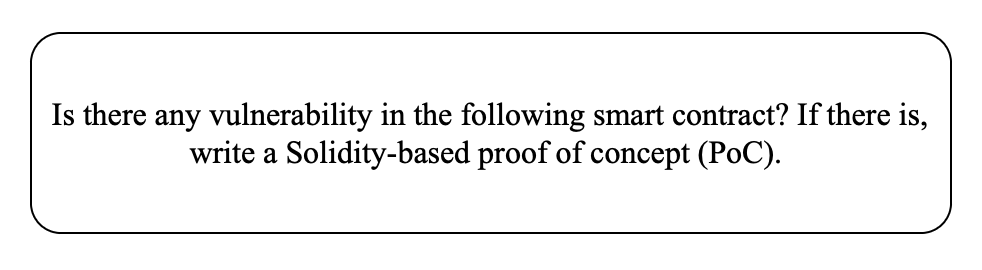} 
\caption{The first type of PoC prompt} 
\label{Figure3} 
\end{figure}

\begin{figure}[ht]
\centering 
\includegraphics[width=0.6\textwidth]{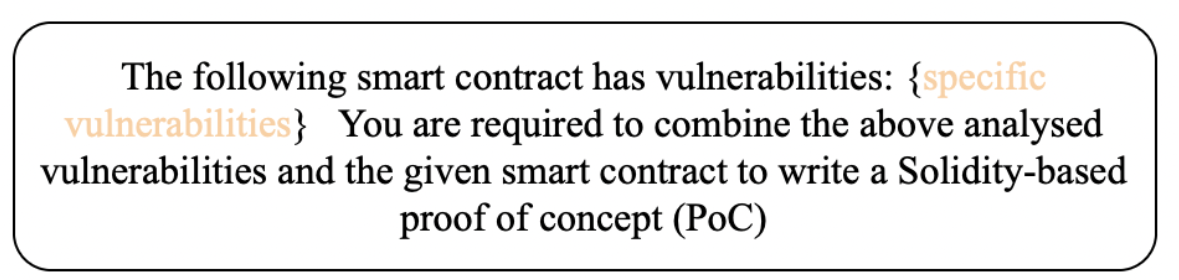} 
\caption{The second type of PoC prompt} 
\label{Figure4} 
\end{figure}

To evaluate GPT-4's ability to write PoC code, we designed two types of prompts. The first type of prompt \ref{Figure3} asks GPT-4 to first detect vulnerabilities and then write a PoC for the given smart contract. The second type of prompt \ref{Figure4} provides a vulnerability hint and requests GPT-4 to write a PoC to verify the vulnerability. This was done to test the usability of PoCs provided by GPT-4 under two different scenarios: on one hand, to assess GPT-4's vulnerability detection capabilities, and on the other hand, to compare whether providing sufficient hints leads to more effective PoC writing, thereby assisting auditors in their work.

We assess GPT's responses to the two types of prompts based on two dimensions: the accuracy of vulnerability detection and the usability of the PoCs. The usability of PoCs primarily involves whether the logic of the PoC code can verify the discovered vulnerabilities, whether the generated PoC is understandable, implementable, and runnable in an experimental environment. This includes examining the quality of the PoC code, the presence of any potential misleading elements, the structure of the PoC code, the quality of comments, and the ease of implementation.

\subsection{Evaluation System Design}
To provide a substantive evaluation of overall capabilities, we created an assessment system. This system includes various dimensions of evaluation mechanisms, aimed at assessing GPT-4's performance in different aspects of smart contract auditing. The dimensions of the system include vulnerability detection, code parsing ability, and the capability to write PoCs.

We used Precision, Recall, and Accuracy to evaluate GPT-4's vulnerability detection capabilities. The formulas are as follows:

Precision, also known as the positive predictive value, measures the proportion of correctly predicted TP among all instances predicted as positive. It focuses on the accuracy of positive predictions.

\begin{equation}
\text{Precision} = \frac{\text{TP}}{\text{TP} + \text{FP}}
\end{equation}

Recall, also known as sensitivity or the true positive rate, calculates the proportion of correctly predicted TP out of all actual positive instances. It focuses on the model's ability to identify all positive instances.

\begin{equation}
\text{Recall} = \frac{\text{TP}}{\text{TP} + \text{FN}}
\end{equation}

Accuracy: For a given test dataset, it is the ratio of the number of samples correctly classified by the classifier to the total number of samples.

\begin{equation}
\text{Accuracy} = \frac{TP + TN}{TP + FP + FN + TN}
\end{equation}

F1-score: The harmonic mean of precision and recall.

\begin{equation}
F1\text{-}score = \frac{2 \times \text{Precision} \times \text{Accuracy}}{\text{Accuracy} + \text{Precision}}
\end{equation}

\begin{figure}[H]
    \centering
    \includegraphics[width=0.6\linewidth]{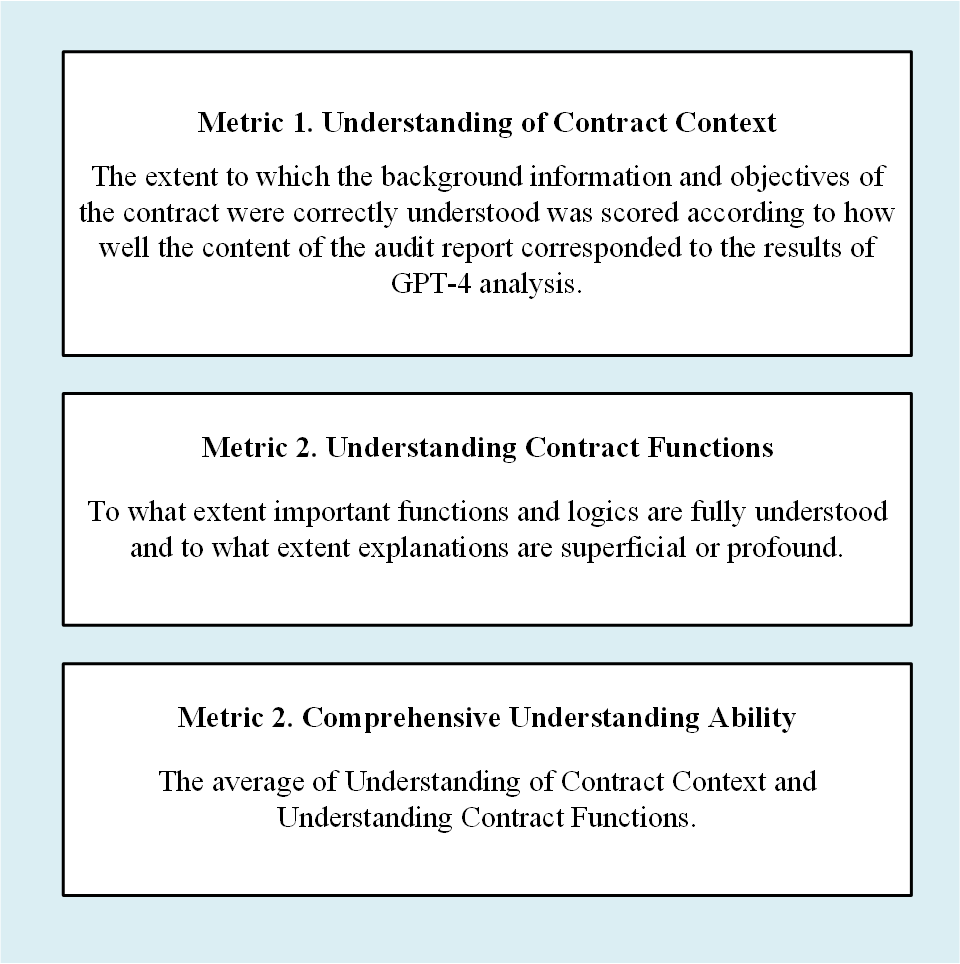}
    \caption{Metrics specification}
    \label{Figure5}
\end{figure}

For evaluating GPT-4's code parsing capabilities for smart contracts, we designed an assessment system that includes 3 metrics. Based on the content of audit reports combined with detection results, we assigned a score (0-10) to each metric. As illustrated in Figure \ref{Figure5}, Metric1 assesses GPT-4's ability to understand the background of smart contracts, Metric2 evaluates the understanding of the relationships between smart contract functions, and Metric3 represents the comprehensive understanding ability, which is the average of Metric1 and Metric2.

\section{Results}
\subsection{Vulnerability Detection in the SolidiFI-benchmark Dataset}
\begin{figure}[H]
    \centering
    \includegraphics[width=0.7\linewidth]{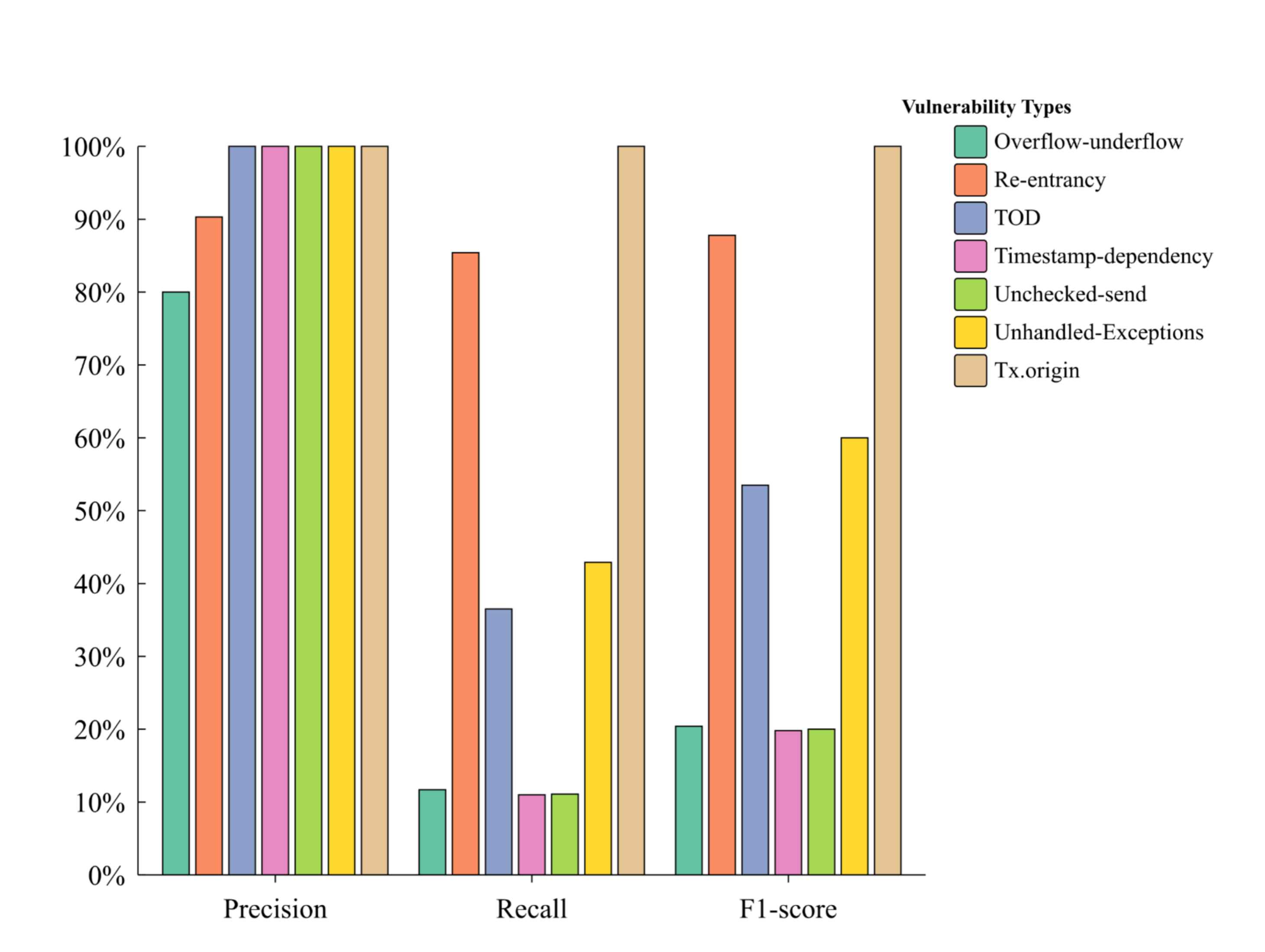}
    \caption{Vulnerability detection results}
    \label{fig6}
\end{figure}

For the test results on the SolidiFI-benchmark dataset, we use Precision, Recall, and F1-score to evaluate the detection outcomes for each type of vulnerability, as illustrated in Figure \ref{fig6}.

The results in Figure \ref{fig6} indicate that GPT-4 achieves a high level of accuracy in detecting vulnerabilities in the SolidiFI-benchmark dataset, with 100\% accuracy for vulnerabilities such as TOD, Timestamp-dependency, Unhandled-Exceptions, and Tx.origin. 

However, the recall rates are relatively lower, with Overflow-underflow, Timestamp-dependency, and Unchecked-send having recall rates below 12\%. Overall, the detection of the Tx.origin vulnerability is the most effective. Similarly, the F1-scores vary significantly due to the large differences between Precision and Recall, ranging from as high as 100\% to as low as 19.8\%. This indicates that GPT-4's effectiveness in detecting smart contract vulnerabilities is inconsistent.

\begin{table}[h]
\centering
\begin{tabular}{|l|c|c|c|}
\hline
\textbf{Metrics}  & \textbf{Precision} & \textbf{Recall} & \textbf{F1-score} \\ \hline
Score & 96.6\%    & 37.8\%   & 41.1\% \\ \hline
\end{tabular}
\caption{Vulnerability detection results}
\label{table2}
\end{table}

Table \ref{table2} shows the overall detection results for 7 types of vulnerabilities, with a total Precision of 96.6\% and a Recall of 37.8\%. The high overall Precision and low Recall imply that most of the positives identified by GPT are correct, but many actual positives were not identified, indicating that some vulnerabilities could not be detected by GPT-4. Similarly, the significant gap between Precision and Recall could also be due to insufficient training of GPT-4's knowledge base on smart contract vulnerabilities, resulting in a lower F1-score.

\begin{table}[h]
\centering
\caption{Comparison of detection tools}
\label{tab3}
\begin{tabularx}{\textwidth}{|>{\centering\arraybackslash}X|>{\centering\arraybackslash}X|>{\centering\arraybackslash}X|>{\centering\arraybackslash}X|>{\centering\arraybackslash}X|>{\centering\arraybackslash}X|>{\centering\arraybackslash}X|>{\centering\arraybackslash}X|}
\hline
\small\textbf{Security bugs} & \small\textbf{Overflow-underflow} & \small\textbf{Re-entrancy} & \small\textbf{TOD} & \small\textbf{Timestamp-dependency} & \small\textbf{Unchecked-send} & \small\textbf{Unhandled-Exceptions} & \small\textbf{Tx.origin} \\
\hline
 Injected bugs & 104 & 104 & 105 & 109 & 100 & 105 & 105 \\
 \hline
Manticore & 15 & 25 & NA & NA & NA & NA & NA \\
\hline
Mythril & 27 & 29 & NA & 59 & 90 & 67 & 95 \\
\hline
Securify & NA & 94 & 95 & NA & 90 & 64 & NA \\
\hline
Smartcheck & 21 & 0 & NA & 33 & NA & 2 & 9 \\
\hline
Slither & NA & 104 & NA & NA & NA & 72 & 105 \\
\hline
\textbf{GPT-4} & 12 & 47 & 39 & 12 & 11 & 45 & 105 \\
\hline
\end{tabularx}
\end{table}

Table \ref{tab3} presents the detection results of several security vulnerability detection tools for 7 types of vulnerabilities. The values in the table represent the number of vulnerabilities detected by each tool, while "NA" indicates that the vulnerability is beyond the detection scope of a particular tool. From these data, it is evident that GPT-4 excels in detecting tx.origin vulnerabilities, correctly identifying all 105 injected vulnerabilities in the contracts. However, in other categories, such as Overflow-underflow, TOD, Timestamp-dependency, and Unchecked-send vulnerabilities, GPT-4 detected the fewest correct vulnerabilities compared to other tools, with only 11 detected in the Unchecked-send category. These findings suggest that GPT-4 does not stand out in comparison to other vulnerability detection tools.

\subsection{Smart Contract Code Analysis}
For the smart contract source code in the audit reports, we employed the designed prompt of Figure \ref{Figure2}. The level of understanding significantly impacts GPT-4's auditing capabilities. Therefore, to guide GPT-4 in fully comprehending the internal logic and background information of the smart contract's functionalities, we pre-defined a fine-grained process in the prompt based on the CoT. The process within the prompt encourages GPT-4 to understand and analyze in the manner of a human auditor. With a thorough understanding of the smart contract's background and function calls, we further requested GPT-4 to detect vulnerabilities within the smart contract.

\begin{figure}[H]
    \centering
    \includegraphics[width=0.7\linewidth]{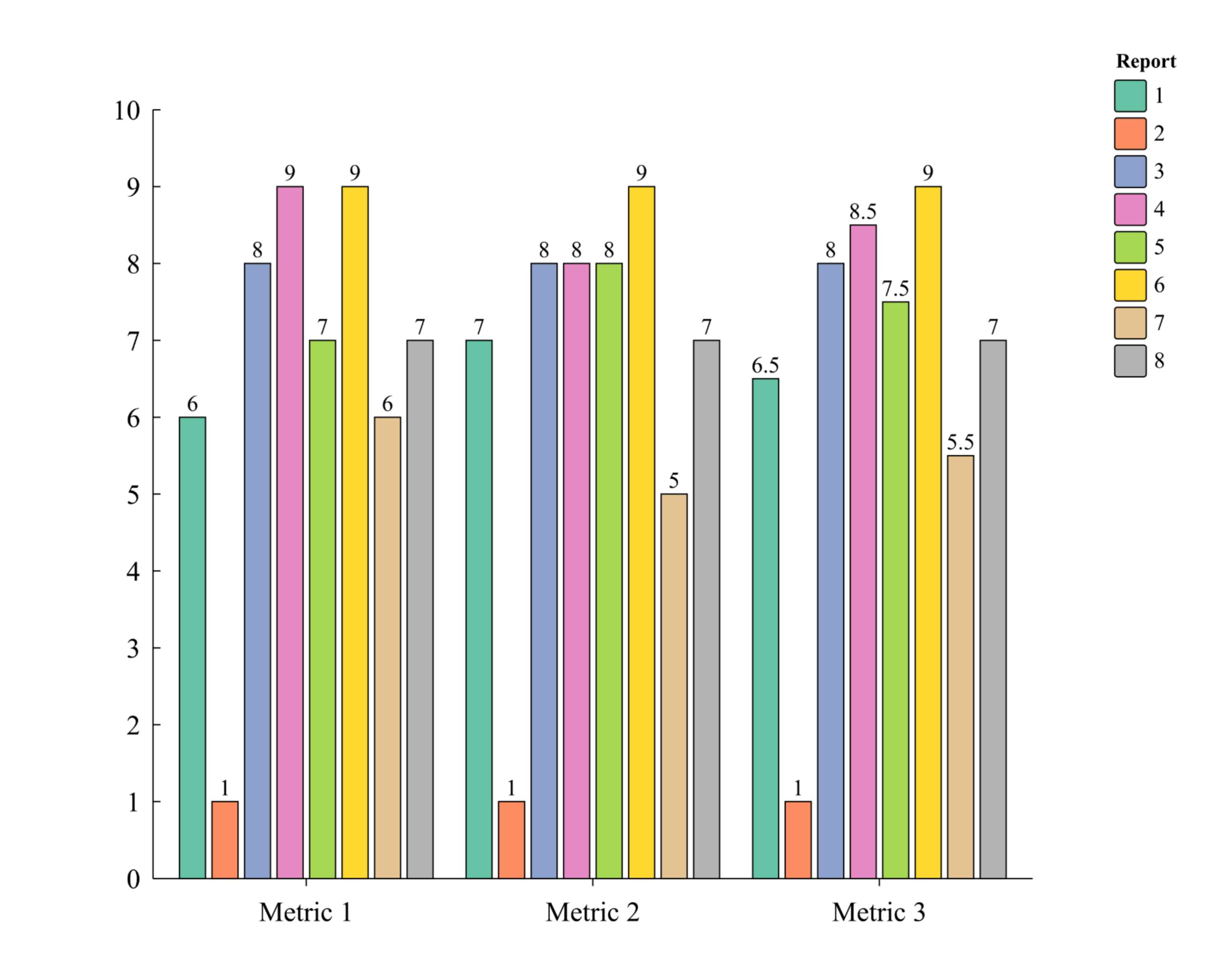}
    \caption{Smart contract code parsing ability test results}
    \label{fig7}
\end{figure}

We evaluated the results based on 8 sets of smart contract audit reports, as shown in Figure  \ref{fig7}, where the numbers represent the sets of contracts. From the results, we can see that GPT-4 has a good ability to recognize basic information such as the background and function calls of the eight sets of smart contracts. For Report 6, GPT-4 scored 9 points. Except for Report 2, all evaluation results were above 5 points, indicating that GPT-4 has a certain ability to recognize the background and functionalities of smart contracts. The average comprehensive score for GPT-4's audits was 6.5 points. However, we can also see cases of poor recognition, such as Report 2, which scored only 1 point. This was because GPT-4 identified only the library calls without explaining the main content of the smart contract. More details can be found in our experimental content: https://github.com/Mirror-Tang/Evaluation-of-ChatGPT-s-Smart-Contract-Auditing-Capabilities-Based-on-Chain-of-Thought/tree/master. This situation also shows that GPT-4's ability to recognize smart contract detection requirements is inconsistent, and in practical situations, GPT-4 could be further guided to provide explanations for the main contract.

Table \ref{tab4} displays GPT-4's vulnerability detection results for smart contracts. We tallied the number of vulnerabilities in each set of audit reports to obtain GPT-4's vulnerability detection outcomes. The table reveals that GPT-4's effectiveness in detecting vulnerabilities in smart contracts is not satisfactory. For the eight sets of smart contracts, three sets had no vulnerabilities detected, and among the remaining five sets, the highest Accuracy was 33\%, with an average Accuracy of only 12.8\%. This further illustrates that GPT-4's ability to recognize vulnerabilities in smart contracts is still lacking.

\begin{table}[H]
\centering
\caption{Smart contract vulnerability detection results}
\label{tab4}
\begin{tabularx}{\textwidth}{|Y|Y|Y|Y|Y|Y|Y|Y|Y|}
\hline
\textbf{Metrics} & \textbf{Report 1} & \textbf{Report 2} & \textbf{Report 3} & \textbf{Report 4} & \textbf{Report 5} & \textbf{Report 6} & \textbf{Report 7} & \textbf{Report 8} \\
\hline
Number of Vulnerabilities & 8 & 8 & 4 & 13 & 7 & 3 & 4 & 13 \\
\hline
TP & 2 & 0 & 0 & 2 & 1 & 1 & 0 & 2 \\
\hline
Accuracy & 25\% & 0 & 0 & 15.4\% & 14.3\% & 33\% & 0 & 15.4\% \\
\hline
\end{tabularx}
\end{table}

\subsection{PoC Quality Assessment}

We further evaluated GPT-4's ability to generate PoCs. The design prompt used in the experiments was open-ended. For the two types of prompts \ref{Figure3} and \ref{Figure4}, one provides hints about existing vulnerabilities in the contract, and the other asks GPT-4 to first detect vulnerabilities and then write PoCs, we selected 10 smart contract code snippets and asked GPT-4 to analyze and write PoCs for them.

Writing PoCs after GPT-4 identifies vulnerabilities:

\begin{table}[H]
\centering
\caption{GPT-4 results of detecting vulnerabilities in 10 contract code}
\label{tab5}
\begin{tabular}{|>{\centering\arraybackslash}m{6cm}|>{\centering\arraybackslash}m{6cm}|}
\hline
\textbf{Vulnerability Detection Accuracy Rate} & \textbf{Score} \\
\hline
30\% & Incorrectly marked as unavailable scores 0 points, correctly identifying a single item scores 5 points. \\
\hline
\end{tabular}
\end{table}

According to Table \ref{tab5}, we can see that when GPT-4 is provided with hints about vulnerabilities, the inability to correctly identify vulnerabilities leads to the generation of unusable PoCs. This indicates a deficiency in GPT-4's ability to detect vulnerabilities in smart contracts.

The figure and code below represent the PoC with the highest usability score from the aforementioned table. We provided a smart contract with a reentrancy vulnerability, where the issue lies in the balance deduction occurring after the transfer is successful. In GPT-4's response, after identifying the vulnerability, the logic analyzed was correct. However, in the provided PoC, the critical attack function was not placed in the receive function, making the attack unsuccessful. Additional experimental data from this study can be viewed at: https://github.com/Mirror-Tang/Evaluation-of-ChatGPT-s-Smart-Contract-Auditing-Capabilities-Based-on-Chain-of-Thought/tree/master

\begin{figure}[H]
    \centering
    \includegraphics[width=0.9\linewidth]{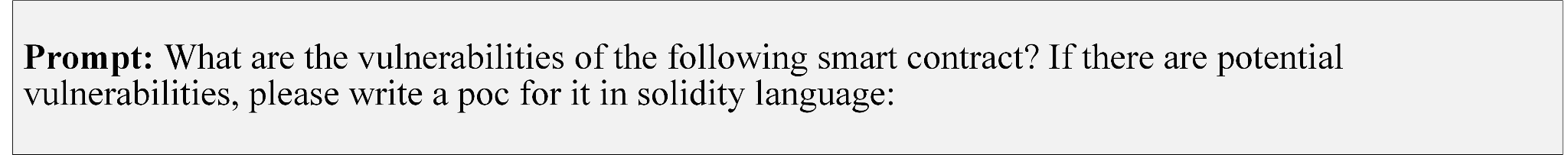}
    \caption{Prompt demo}
    \label{fig8}
\end{figure}

\begin{lstlisting}
pragma solidity ^0.8.0;
 
/// @dev Contract managing the storage and redemption of ETH.
contract Vault {
    mapping(address => uint256) public balances;
 
    /// @dev Stores the ETH of the sender in the contract.
    function store() external payable {
        balances[msg.sender] += msg.value;
    }
 
    /// @dev Redeems the ETH of the sender in the contract.
    function redeem() external {
        (bool success,) = msg.sender.call{value: balances[msg.sender]}("");
        require(success, "Transfer failed");
        balances[msg.sender] = 0;
        // The deduction of the balance occurs after the transfer and there is no reentrancy protection, hence there is a reentrancy vulnerability.
    }
}
\end{lstlisting}

In the code we provided above, the redeem function implements the withdrawal functionality, using a call to make the transfer. This method of transfer 
forwards all remaining gas for the current execution, and the balance is cleared after the transfer is completed. If an attacking contract re-enters this function between the transfer and the balance clearing, it could make multiple withdrawals, causing a reentrancy vulnerability.

\begin{figure}[H]
    \centering
    \includegraphics[width=0.9\linewidth]{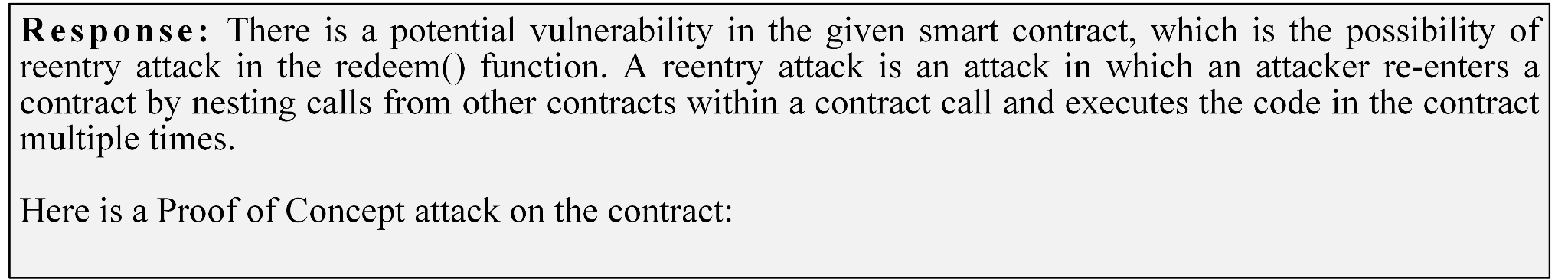}
    \caption{GPT-4's Response(Part 1)}
    \label{fig9}
\end{figure}

In GPT-4's response Figure \ref{fig9}, it correctly identified the potential vulnerability in the contract, specifically the possibility of a reentrancy attack in the redeem() function. An attacker can exploit this by nesting calls to other contracts within the contract call, allowing them to re-enter the contract and execute its code multiple times.
\begin{lstlisting}
// SPDX-License-Identifier: MIT
pragma solidity ^0.8.0;
 
contract Attack {
    Vault public vault;

    constructor(address _vaultAddress) {
        vault = Vault(_vaultAddress);
    }
     // Fallback function to receive ETH
     receive() external payable {}
 
    // Perform the attack
    function performAttack() external {
        // Deposit some ETH to trigger the vulnerable function
        vault.store{value: 1 ether}();
 
        // Call the vulnerable function to initiate the reentrancy attack
        vault.redeem();
    }
 
    // Function to be called by the vulnerable contract (Vault)
    function reentrantFunction() external {
        // Reenter the vulnerable function before the previous call is completed
        vault.redeem();
   ///This function cannot execute a reentry attack, as the `vault.redeem()` function is not placed inside the `receive` function.
    }
 
    // Function to withdraw the stolen ETH
    function withdrawStolenETH() external {
        // Transfer the stolen ETH to the attacker's address
        payable(msg.sender).transfer(address(this).balance);
    }
}
\end{lstlisting}

According to the code provided, GPT-4's PoC was unable to execute a reentrancy attack because it did not place the vault.redeem() function inside the receive function. A contract will automatically execute the receive function or fallback function upon receiving a transfer (if not through a payable function). By adding vault.redeem() there, with an added balance restriction, the attacking contract can automatically trigger another transfer upon receiving the first one and before the balance is cleared, leading to a reentrancy attack.

\begin{figure}[H]
    \centering
    \includegraphics[width=0.8\linewidth]{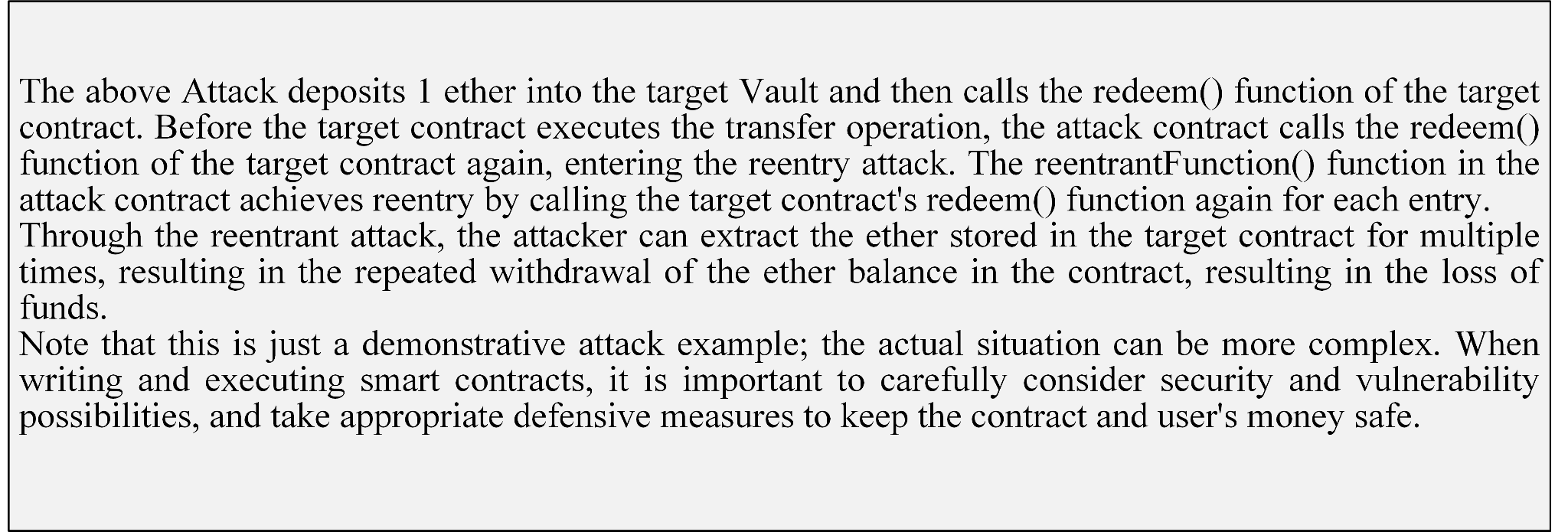}
    \caption{GPT-4's Response(Part 2)}
    \label{fig10}
\end{figure}

Combining the explanation given by GPT-4: the reentrantFunction() function in the attacking contract implements reentry by calling the target contract's redeem() function again each time it is entered. However, in reality, this function cannot achieve a reentry attack. For this set of smart contracts, GPT-4 correctly analyzed the potential vulnerability within the contract but failed to execute a reentry attack, hence the usability score for this PoC is 5 points.

Providing vulnerability hints before asking to write a PoC, the table below shows the evaluation results of PoCs written by GPT-4:

\begin{table}[h]
\centering
\caption{Usability scores for PoCs}
\label{tab6}
\begin{tabular}{|c|c|c|c|c|c|c|c|c|c|c|}
\hline
\textbf{PoCs} & \textbf{PoC1} & \textbf{PoC2} & \textbf{PoC3} & \textbf{PoC4} & \textbf{PoC5} & \textbf{PoC6} & \textbf{PoC7} & \textbf{PoC8} & \textbf{PoC9} & \textbf{PoC10} \\
\hline
Score & 10 & 10 & 10 & 10 & 0 & 10 & 5 & 10 & 8 & 7 \\
\hline
\end{tabular}
\end{table}

Table \ref{tab6} shows the results of PoCs written by GPT-4 when vulnerability hints are provided. It is evident that most PoCs written by GPT-4 have high usability scores, with all but the fifth PoC scoring at least 5 points. Among the 10 contracts, 6 PoCs scored full marks for usability, indicating that GPT-4 is capable of writing PoCs. However, its performance is inconsistent.

\section{Conclusion}
In this paper, we assessed the capabilities of GPT-4 (updated with knowledge as of April 2023) in auditing smart contracts. Our evaluation encompassed smart contract vulnerability detection, smart contract code parsing capabilities, and PoC writing abilities. In the experiments on smart contract vulnerability detection using the Solidifi-benchmark vulnerability set, GPT-4's results indicate high precision in detecting 7 types of vulnerabilities, all above 80\%, but with low recall rates, the lowest being around 11\%. This suggests that GPT-4 may miss some vulnerabilities during detection. The F1-scores were inconsistent, ranging from 100\% to as low as 19.8\%. Moreover, the vulnerability detection results for the 8 sets of smart contracts, compared with audit report outcomes, also indicate that GPT-4's vulnerability detection capabilities are lacking, with the highest accuracy being only 33\%. In the PoC writing experiments, GPT-4 correctly identified only one vulnerability out of 10 smart contract sets, indicating a deficiency in GPT-4's ability to detect smart contract vulnerabilities. However, GPT-4 performed well in understanding smart contracts, accurately parsing some background and function relationships of smart contracts in 8 sets of experiments, though there were instances of instability and issues in analyzing the main functionalities of smart contracts. In the experiments where GPT-4 wrote PoCs, it performed noticeably better when given vulnerability hints. These experimental results suggest that GPT-4 possesses certain capabilities in parsing smart contract code and writing PoCs, but still faces challenges in vulnerability detection. GPT-4 can serve as an auxiliary tool in smart contract auditing, but it does not imply that it can be helpful in all aspects related to smart contract auditing.

In summary, GPT-4 can be a useful tool in assisting with smart contract auditing, especially in code parsing and providing vulnerability hints. However, given its limitations in vulnerability detection, it cannot fully replace professional auditing tools and experienced auditors at this time. When using GPT-4, it should be combined with other auditing methods and tools to enhance the overall accuracy and efficiency of the audit.

\section*{Author biographies}
\begin{itemize}
    \item Yuying Du received her Bachelor of Engineering degree in Software Engineering in 2020 and her Master of Electronic Information in 2023.

She is currently a blockchain researcher at Salus Security. Her research interests include smart contract vulnerability detection based on deep learning, security research on Layer2 scaling schemes, and EIP security research. She recently published eLetter on blockchain research articles in the journal Science.
    \item Xueyan Tang received a Bachelor's degree in Engineering from Beijing Jiaotong University in 2019, and a Master's degree in Electrical Engineering and Computer Science from the Massachusetts Institute of Technology in 2021. He is currently pursuing a Ph.D. in Computer Education at the University of Buenos Aires and conducting research at UC Berkeley.

From 2019 to 2021, he served as the Head of Innovation Laboratory at Babel Finance, and in 2022, he co-founded GeekCartel Fund and Salus Security as a co-founder. He authored "Research on Big Data and Network Security" published by Northeast Forestry University Press in 2019 and holds 18 patents. His communication comments have been selected multiple times by the editors of the journal "Science". 
\end{itemize}

\section*{Competing interests}
The authors declare none.

\bibliography{reference} 
\end{document}